# John Equation Constraints for the 3D X-ray Transform under a Cylindrical-Spherical Mixed Parameterization: Theoretical Derivation, Experimental Validation, and Application Analysis

Shaojie Tang[1], Zhiwei Qiao[2], Xuanqin Mou[3]

1. School of Artificial Intelligence and School of Automation, Xi'an University of Posts and Telecommunications, Xi'an, Shaanxi 710121, China

2. School of Computer and Information Technology, Shanxi University, Taiyuan, Shanxi 030006, China

3. School of Electronic and Information Engineering, Xi'an Jiaotong University, Xi'an, Shaanxi 710121, China

**Abstract:** The John equation serves as the mathematical foundation of the X-ray transform, describing the intrinsic compatibility conditions that projection data must satisfy. In this paper, within three-dimensional (3D) Euclidean space, an innovative mixed parameterization scheme is adopted: the source point is represented using cylindrical coordinates $a = (s\cos\theta, s\sin\theta, z_0)$, and the ray direction is represented using spherical coordinates $d = \rho(-\cos\beta\sin\alpha, \cos\beta\cos\alpha, \sin\beta)$. The specific form of the John equation under this geometric parameterization is systematically derived. Through detailed partial differential operator transformations, application of -1 homogeneity, and algebraic simplification, a complete system of constraint equations is obtained. In particular, under the special configurations where the ray direction is perpendicular to the radial direction of the source point in the horizontal plane (i.e., the so-called alignment condition: $\alpha = \theta$) and the ray has no tilt $(\beta = 0)$, the constraint equations simplify to differential relations with clear physical meanings. This paper not only establishes a bridge between abstract mathematical theory and concrete imaging geometry, but also provides rigorous mathematical tools for data consistency verification, geometric parameter calibration, and incomplete-data reconstruction in 3D Computed Tomography (CT) systems. The research results are of great significance for advancing the mathematical theory and practical applications of CT imaging.

## 1. Introduction

Computed Tomography (CT) technology's mathematical foundation is the Radon transform [1] and its generalization—the X-ray transform [2]. In CT imaging, projection data must satisfy certain internal compatibility conditions, also known as projection data consistency conditions [3-9]. However, most projection data consistency conditions involve integral operations, which are very unfavorable for practical applications. The John equation, originally proposed by Fritz John in 1938 [3], has a very favorable partial differential equation (PDE) form and constitutes one of the important foundations of X-ray transform theory, providing a mathematical criterion for projection data consistency.

However, the original John equation is expressed in Cartesian coordinates, with an abstract form and lacking intuitive interpretation for specific scanning geometries [3]. In practical CT imaging systems, scanning trajectories usually possess specific geometric symmetries, such as circular, helical, or linear trajectories [2,10-15,24,25]. How to transform the abstract John equation into practical constraint equations for specific scanning geometries has always been an important subject in CT mathematical theory [9,23].

This study addresses this problem. An innovative mixed parameterization scheme is proposed: the source point position is represented using cylindrical coordinates, which has a good correspondence with the 2D Radon transform and can naturally describe the circular motion of the source point around the rotation axis; the ray direction is represented using a specific form of spherical coordinates, which reflects both the spatial orientation of the ray and contains the ray length information. This parameterization scheme has clear physical meanings: $s$ represents the radial distance from the source point to the rotation axis, $\theta$ represents the rotation angle of the source point, and $z_0$ represents the axial position of the source point; $\rho$ represents the ray length, i.e., the distance from the source point to the detector point, $\alpha$ represents the azimuthal angle of the ray direction in the horizontal plane, defining the position of the detector unit in the detector array, and $\beta$ represents the tilt angle of the ray direction relative to the horizontal plane. This parameterization can cover the scanning geometries of most practical CT systems, including cone-beam CT, helical CT, and multi-detector CT [2,10-15,24,25].

The main contributions of this paper include: 1. Establishing a complete mathematical derivation framework for the John equation under mixed parameterization; 2. Obtaining the specific constrained PDE system; 3. Analyzing the simplified forms under special geometric configurations and their physical meanings; 4. Discussing the potential applications of the constraint equations in CT imaging.

## 2. Theoretical Foundations

### 2.1 Mathematical Definition of the X-ray Transform

In $\mathbb{R}^3$, the X-ray transform maps an object function $u(x)$: $\mathbb{R}^3 \to \mathbb{R}$ to its line integral along a straight line [2]. For a given source point $a = (a_1, a_2, a_3) \in \mathbb{R}^3$ and detector point $b = (b_1, b_2, b_3) \in \mathbb{R}^3$, the X-ray transform is defined as [2,24]:

$$f(a, b) = \int u(a + t(b - a))\, dt \tag{2.1}$$

where $t$ is the parameter along the ray direction, and $u(x)$ usually represents the distribution of the object's linear attenuation coefficient. This definition assumes that the ray emanates from point a and propagates along the straight line direction $b - a$. The function $f(a, b)$ represents the projection value parameterized by $a$ and $b$; note that this function is not yet strictly the X-ray transform and requires subsequent adjustment.

## 2.2 Basic Form of the John Equation

Fritz John proved that the X-ray transform function $f(a, b)$ satisfies the following second-order PDE [3]:

$$\partial^2 f/\partial a_i \partial b_j = \partial^2 f/\partial a_j \partial b_i, \quad \forall\, i, j = 1, 2, 3 \tag{2.2}$$

which contains 9 component equations, but due to symmetry, only 3 are independent. Usually, the three equations corresponding to $(i, j) = (1, 2), (1, 3), (2, 3)$ are taken as the independent constraints.

The physical meaning of this equation is that it reflects the property that mixed partial derivatives in the X-ray transform are independent of the order of differentiation. This symmetry stems from the essence of the ray integral—regardless of whether one differentiates with respect to the source point coordinates first or the detector point coordinates first, the final result should be the same. From the perspective of data compatibility, the John equation provides the internal consistency conditions that projection data must satisfy: any set of valid projection data must satisfy this system of equations; conversely, data that does not satisfy this system of equations must contain inconsistencies, possibly caused by measurement noise, geometric errors, or physical effects (such as scattering).

## 2.3 John Equation Based on Ray Direction

Introducing the ray direction vector $d = b - a = (d_1, d_2, d_3)$, define the reduced function $g(a, d) = f(a, a + d) = f(a, b)$. Then the X-ray transform (2.1) can be rewritten as:

$$g(a,d) = \int u(a + td)\, dt \tag{2.3}$$

Under this parameterization, the John equation needs to be transformed from the $(a, b)$ coordinate system to the $(a, d)$ coordinate system.

From Appendix A, the form after coordinate system transformation is obtained:

$$\partial^2 g/\partial a_i \partial d_j = \partial^2 g/\partial a_j \partial d_i, \quad \forall\, i, j = 1, 2, 3 \tag{2.4}$$

That is, the John equation based on ray direction, whose form is similar to (2.2), but the partial derivative variables change from $b$ to $d$. This form will be more convenient for subsequent parameterization transformations.

## 3. Mixed Parameterization Scheme

### 3.1 Parameterization Motivation and Geometric Interpretation

In practical CT imaging systems, the source point and ray direction usually follow specific geometric constraints. For example, in third-generation CT, the X-ray source and detector perform synchronous circular motion around the rotation center; in helical CT, this circular motion is superimposed with axial translational motion. To describe this geometry more naturally, the following mixed parameterization scheme is proposed, corresponding to the so-called Wedge imaging geometry, but with a good correspondence to the 2D Radon transform, and capable of naturally describing the circular motion of the source point around the rotation axis, as shown in Fig. 1.

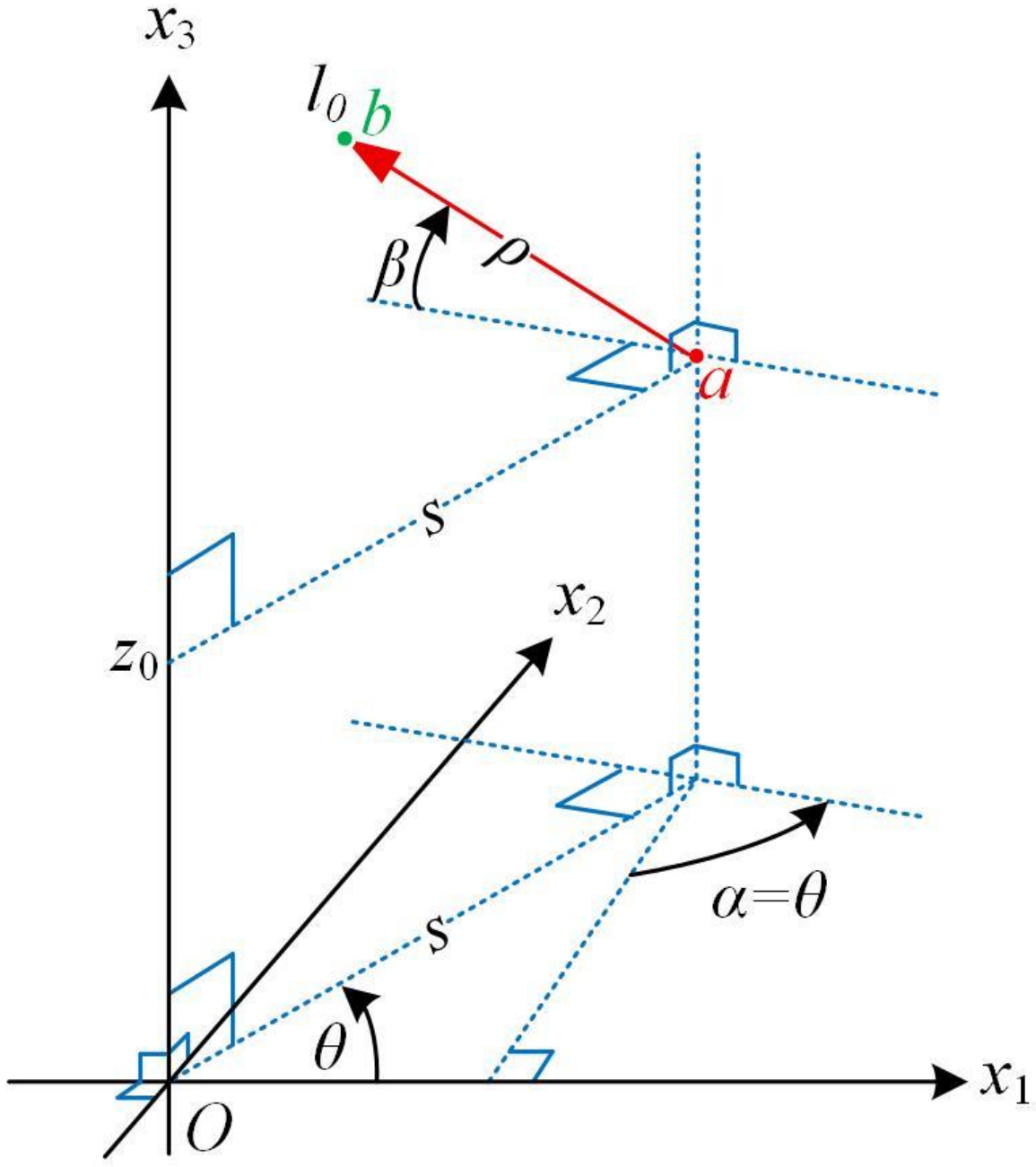

Fig. 1 Imaging geometry with the mixed parameterization scheme

#### 3.1.1 Cylindrical Coordinate Parameterization of the Source Point

The source point $a$ is represented using cylindrical coordinates:

$$a_1 = s \cos \theta;\ a_2 = s \sin \theta;\ a_3 = z_0 \tag{3.1}$$

where $s \geq 0$ is the radial distance from the source point to the z-axis (rotation axis). $\theta \in [0, 2\pi)$ is the azimuthal angle of the source point in the rotation plane. $z_0 \in \mathbb{R}$: the axial coordinate of the source point.

#### 3.1.2 Spherical Coordinate Parameterization of the Ray Direction

The ray direction vector $d = (d_1, d_2, d_3)$ is parameterized as follows:

$$d_1 = -\rho \cos \beta \sin \alpha; \quad d_2 = \rho \cos \beta \cos \alpha; \quad d_3 = \rho \sin \beta \tag{3.2}$$

where $\rho > 0$ is the ray length, i.e., the distance from the source point to the detector point. $\alpha \in [0, 2\pi)$ is the azimuthal angle of the ray direction in the horizontal plane, defining the position of the detector unit in the detector array. $\beta \in (-\pi/2, \pi/2)$: the tilt angle of the ray direction relative to the horizontal plane. It is particularly noteworthy that when $\alpha = \theta$, the ray direction is perpendicular to the radial direction of the source point in the horizontal plane; when $\beta = 0$, the ray lies in the horizontal plane.

### 3.2 Definition of the Reduced Function

To simplify subsequent derivations, we define the reduced function $G$, expressing $g(a, d)$ with new parameters:

$$G(s, \theta, z_0, \rho, \alpha, \beta) = g(a(s, \theta, z_0), d(\rho, \alpha, \beta)) \tag{3.3}$$

where $G$ contains all six geometric parameters: three source point parameters $(s, \theta, z_0)$ and three ray direction parameters $(\rho, \alpha, \beta)$. In CT imaging, for a fixed object $u(x)$, the function $G$ depends on the scanning geometry and ray direction.

## 4. Coordinate Transformation of Partial Differential Operators

### 4.1 Partial Derivative Transformation with Respect to Source Point Coordinates

By the chain rule, the partial derivative with respect to the source point coordinate $a_i$ can be expressed as:

$$\partial/\partial a_i = (\partial s/\partial a_i)(\partial/\partial s) + (\partial\theta/\partial a_i)(\partial/\partial\theta) + (\partial z_0/\partial a_i)(\partial/\partial z_0) \tag{4.1}$$

According to the cylindrical coordinate relations (3.1), see Appendix B for the calculation of each partial derivative:

$$\partial/\partial a_1 = \cos\theta\, \partial/\partial s - (\sin\theta/s)\, \partial/\partial\theta; \quad \partial/\partial a_2 = \sin\theta\, \partial/\partial s + (\cos\theta/s)\, \partial/\partial\theta; \quad \partial/\partial a_3 = \partial/\partial z_0 \tag{4.2}$$

### 4.2 Partial Derivative Transformation with Respect to Ray Direction Coordinates

By the chain rule, the partial derivative with respect to the ray direction coordinate $d_i$ can be expressed as:

$$\partial/\partial d_i = (\partial\rho/\partial d_i)(\partial/\partial\rho) + (\partial\alpha/\partial d_i)(\partial/\partial\alpha) + (\partial\beta/\partial d_i)(\partial/\partial\beta) \tag{4.3}$$

According to the cylindrical coordinate relations (3.2), see Appendix C for the calculation of each partial derivative:

$$\partial/\partial d_1 = -\cos\beta \sin\alpha\, \partial/\partial\rho - \cos\alpha/(\rho\cos\beta)\, \partial/\partial\alpha + \sin\beta\sin\alpha/\rho\, \partial/\partial\beta;$$

$$\partial/\partial d_2 = \cos\beta \cos\alpha\ \partial/\partial\rho - \sin\alpha/(\rho \cos\beta)\ \partial/\partial\alpha - \sin\beta \cos\alpha/\rho\ \partial/\partial\beta;$$

$$\partial/\partial d_3 = \sin\beta\ \partial/\partial\rho + \cos\beta/\rho\ \partial/\partial\beta \tag{4.4}$$

The partial derivative operators in Cartesian coordinates have been completely transformed to the new parameter system, laying the foundation for the subsequent derivation of the specific form of the John equation.

## 5. Expansion and Simplification of the John Equation

### 5.1 -1 Homogeneity

The X-ray transform has an important -1 homogeneity property: for any positive scalar $\lambda > 0$, we have

$$g(a, \lambda d) = (1/\lambda)\ g(a, d) \tag{5.1}$$

which stems from the definition of the line integral: scaling the ray direction vector proportionally is equivalent to changing the length of the integration path, and the line integral value is inversely proportional to the path length. In the reduced function $G$, the -1 homogeneity is manifested as:

$$G(s, \theta, z_0, \lambda\rho, \alpha, \beta) = (1/\lambda)\ G(s, \theta, z_0, \rho, \alpha, \beta) \tag{5.2}$$

Differentiating with respect to $\lambda$ and setting $\lambda = 1$ yields:

$$\partial G/\partial\rho = -(1/\rho)\ G \tag{5.3}$$

Using this relation, all partial derivatives with respect to $\rho$ can be completely eliminated.

### 5.2 Expansion of Independent John Equations

The John equation (2.4) contains three independent equations, corresponding to $(i, j) = (1, 2), (1, 3), (2, 3)$ respectively.

**Equation J12: $\partial^2 g/\partial a_1\partial d_2 = \partial^2 g/\partial a_2\partial d_1$**

Left-hand side:

$$\partial/\partial a_1\ (\partial g/\partial d_2) = (\cos\theta\ \partial/\partial s - (\sin\theta/s)\ \partial/\partial\theta)\ [\cos\beta \cos\alpha\ \partial G/\partial\rho$$

$$- \sin\alpha/(\rho \cos\beta)\ \partial G/\partial\alpha - \sin\beta \cos\alpha/\rho\ \partial G/\partial\beta] \tag{5.4}$$

Right-hand side:

$$\partial/\partial a_2\ (\partial g/\partial d_1) = (\sin\theta\ \partial/\partial s + (\cos\theta/s)\ \partial/\partial\theta)\ [-\cos\beta \sin\alpha\ \partial G/\partial\rho$$

$$- \cos\alpha/(\rho \cos\beta)\ \partial G/\partial\alpha + \sin\beta \sin\alpha/\rho\ \partial G/\partial\beta] \tag{5.5}$$

**Equation J13:** $\partial^2 g/\partial a_1\partial d_3 = \partial^2 g/\partial a_3\partial d_1$
Left-hand side:

$$\partial^2 g/\partial a_1\partial d_3 = (\cos\theta\ \partial/\partial s - (\sin\theta/s)\ \partial/\partial\theta)\ [\sin\beta\ \partial G/\partial\rho + \cos\beta/\rho\ \partial G/\partial\beta] \quad (5.6)$$

Right-hand side:

$$\partial^2 g/\partial a_3\partial d_1 = \partial/\partial z_0\ [-\cos\beta\sin\alpha\ \partial G/\partial\rho - \cos\alpha/(\rho\cos\beta)\ \partial G/\partial\alpha$$

$$+ \sin\beta\sin\alpha/\rho\ \partial G/\partial\beta] \quad (5.7)$$

**Equation J23:** $\partial^2 g/\partial a_2\partial d_3 = \partial^2 g/\partial a_3\partial d_2$
Left-hand side:

$$\partial^2 g/\partial a_2\partial d_3 = (\sin\theta\ \partial/\partial s + (\cos\theta/s)\ \partial/\partial\theta)\ [\sin\beta\ \partial G/\partial\rho + \cos\beta/\rho\ \partial G/\partial\beta] \quad (5.8)$$

Right-hand side:

$$\partial^2 g/\partial a_3\partial d_2 = \partial/\partial z_0\ [\cos\beta\cos\alpha\ \partial G/\partial\rho - \sin\alpha/(\rho\cos\beta)\ \partial G/\partial\alpha$$

$$- \sin\beta\cos\alpha/\rho\ \partial G/\partial\beta] \quad (5.9)$$

## 5.3 Substituting -1 Homogeneity to Eliminate $\rho$

Substituting the -1 homogeneity into the above equations, all partial derivatives with respect to $\rho$ are eliminated using $\partial G/\partial\rho = -(1/\rho)\ G$. After detailed algebraic operations and eliminating $\rho$ in the denominators, the simplified equations are obtained.

**Simplified form of equation J12:**

$$\partial^2 G/\partial\theta\partial\alpha = s\cos^2\beta\ \partial G/\partial s + \cos^2\beta\tan(\alpha - \theta)\ \partial G/\partial\theta + s\tan(\alpha - \theta)\ \partial^2 G/\partial s\partial\alpha$$

$$+ s\cos\beta\sin\beta\ \partial^2 G/\partial s\partial\beta + \cos\beta\sin\beta\tan(\alpha - \theta)\ \partial^2 G/\partial\theta\partial\beta \quad (5.10)$$

**Simplified form of equation J13:**

$$-\sin\beta\ (\cos\theta\ \partial G/\partial s - (\sin\theta/s)\ \partial G/\partial\theta) + \cos\beta\ (\cos\theta\ \partial^2 G/\partial s\partial\beta - (\sin\theta/s)\ \partial^2 G/\partial\theta\partial\beta)$$

$$= \cos\beta\sin\alpha\ \partial G/\partial z_0 - (\cos\alpha/\cos\beta)\ \partial^2 G/\partial z_0\partial\alpha + \sin\beta\sin\alpha\ \partial^2 G/\partial z_0\partial\beta \quad (5.11)$$

**Simplified form of equation J23:**

$$-\sin\beta\ (\sin\theta\ \partial G/\partial s + (\cos\theta/s)\ \partial G/\partial\theta) + \cos\beta\ (\sin\theta\ \partial^2 G/\partial s\partial\beta + (\cos\theta/s)\ \partial^2 G/\partial\theta\partial\beta)$$

$$= -\cos\beta\cos\alpha\ \partial G/\partial z_0 - (\sin\alpha/\cos\beta)\ \partial^2 G/\partial z_0\partial\alpha - \sin\beta\cos\alpha\ \partial^2 G/\partial z_0\partial\beta \quad (5.12)$$

**Combining equations J13 and J23:**

$$-\tan\beta\ \partial G/\partial s + \partial^2 G/\partial s\partial\beta = \sin(\alpha - \theta)\ \partial G/\partial z_0$$

$$- (1/\cos^2\beta)\cos(\alpha - \theta)\ \partial^2 G/\partial z_0\partial\alpha + \tan\beta\sin(\alpha - \theta)\ \partial^2 G/\partial z_0\partial\beta \quad (5.13)$$

$$-(\tan\beta/s)\ \partial G/\partial\theta + (1/s)\ \partial^2 G/\partial\theta\partial\beta = -\cos(\alpha - \theta)\ \partial G/\partial z_0$$

$$- (1/\cos^2\beta)\ \sin(\alpha - \theta)\ \partial^2 G/\partial z_0\partial\alpha - \tan\beta\ \cos(\alpha - \theta)\ \partial^2 G/\partial z_0\partial\beta \quad (5.14)$$

The above formulas constitute the core constraints of the John equation under special geometric configurations, holding when the reduced function $G$ satisfies -1 homogeneity. The above formulas no longer explicitly contain $\rho$. They explicitly express the constraint relations of the projection data $G$ under variations of different geometric parameters.

## 6. Analysis of Special Geometric Configurations

### 6.1 Alignment Condition: $\alpha = \theta$

In many CT scanning geometries, there is a specific relationship between the azimuthal angle $\alpha$ of the ray direction and the azimuthal angle $\theta$ of the source point. The simplest case is when the two are equal: $\alpha = \theta$. This corresponds to the configuration where the ray direction is perpendicular to the radial direction of the source point in the horizontal plane.

Equation J12 at $\alpha = \theta$:

$$\partial^2 G/\partial\theta\partial\alpha = s\cos^2\beta\ \partial G/\partial s + s\cos\beta\sin\beta\ \partial^2 G/\partial s\partial\beta \quad (6.1)$$

Equation J13 at $\alpha = \theta$:

$$-\sin\beta\ (\cos\theta\ \partial G/\partial s - (\sin\theta/s)\ \partial G/\partial\theta) + \cos\beta\ (\cos\theta\ \partial^2 G/\partial s\partial\beta - (\sin\theta/s)\ \partial^2 G/\partial\theta\partial\beta)$$

$$= \cos\beta\sin\theta\ \partial G/\partial z_0 - (\cos\theta/\cos\beta)\ \partial^2 G/\partial z_0\partial\alpha + \sin\beta\sin\theta\ \partial^2 G/\partial z_0\partial\beta \quad (6.2)$$

Equation J23 at $\alpha = \theta$:

$$-\sin\beta\ (\sin\theta\ \partial G/\partial s + (\cos\theta/s)\ \partial G/\partial\theta) + \cos\beta\ (\sin\theta\ \partial^2 G/\partial s\partial\beta + (\cos\theta/s)\ \partial^2 G/\partial\theta\partial\beta)$$

$$= -\cos\beta\cos\theta\ \partial G/\partial z_0 - (\sin\theta/\cos\beta)\ \partial^2 G/\partial z_0\partial\alpha - \sin\beta\cos\theta\ \partial^2 G/\partial z_0\partial\beta \quad (6.3)$$

Combining equations J13 and J23 at $\alpha = \theta$:

$$-\cos\beta\sin\beta\ \partial G/\partial s + \cos^2\beta\ \partial^2 G/\partial s\partial\beta = -\partial^2 G/\partial z_0\partial\alpha \quad (6.4)$$

$$-\tan\beta\ \partial G/\partial\theta + \partial^2 G/\partial\theta\partial\beta = -s\ \partial G/\partial z_0 - s\tan\beta\ \partial^2 G/\partial z_0\partial\beta \quad (6.5)$$

The above formulas constitute the core constraints of the John equation under special geometric configurations, independent of the specific value of $\theta$. For the relationship between the above formulas and the translation invariance condition, see Appendix D; for numerical validation, see Appendices E-H.

### 6.2 Untilted Alignment Case: $\alpha = \theta, \beta = 0$

Another important special case is the untilted alignment, i.e., $\alpha = \theta, \beta = 0$, corresponding to the situation where the ray lies in the horizontal plane.

Equation J12 at $\alpha = \theta, \beta = 0$:

$$\partial^2 G/\partial\theta\partial\alpha = s\ \partial G/\partial s \tag{6.6}$$

Combining equations J13 and J23 at $\alpha = \theta, \beta = 0$:

$$\partial^2 G/\partial s\partial\beta = -\partial^2 G/\partial z_0\partial\alpha \tag{6.7}$$

$$\partial^2 G/\partial\theta\partial\beta = -s\ \partial G/\partial z_0 \tag{6.8}$$

## 7. Physical Meaning and Geometric Interpretation

Although formulas (6.6)-(6.8) have concise forms, they contain rich physical meanings. Below is a detailed explanation of the geometric meaning of each equation.

### 7.1 Formula (6.6): $\partial^2 G/\partial\theta\partial\alpha = s\ \partial G/\partial s$

This formula relates three geometric parameters: the source point azimuthal angle $\theta$, the ray direction azimuthal angle $\alpha$, and the source point radial distance $s$. Left-hand side: $\partial^2 G/\partial\theta\partial\alpha$ represents the mixed rate of change of the projection value $G$ with respect to the source point azimuthal angle $\theta$ and the ray direction azimuthal angle $\alpha$. In CT imaging, this reflects the rate of change of the projection value when both the source point and the ray direction rotate slightly simultaneously. Right-hand side: $s\partial G/\partial s$ represents the rate of change of the projection value with respect to the source point radial distance $s$, multiplied by $s$. $\partial G/\partial s$ describes the change in the projection value when the source point moves along the radial direction.

Physical meaning: The variation characteristics of the projection data in the azimuthal direction are directly related to its variation characteristics in the radial direction. Specifically, the mixed sensitivity of the projection data to $\theta$ and $\alpha$ equals its sensitivity to $s$ multiplied by $s$. This reflects the coupling relationship between radial distance variation and angular variation in CT systems, as shown in Fig. 2. In practical applications, if the variation law of the projection data with the source point radial distance is known (for example, by changing the source-axis distance), the variation characteristics of the data in the angular direction can be predicted, and vice versa. It is particularly noteworthy that formula (6.6) has a very important intrinsic connection with the recent results of Mou et al. [16].

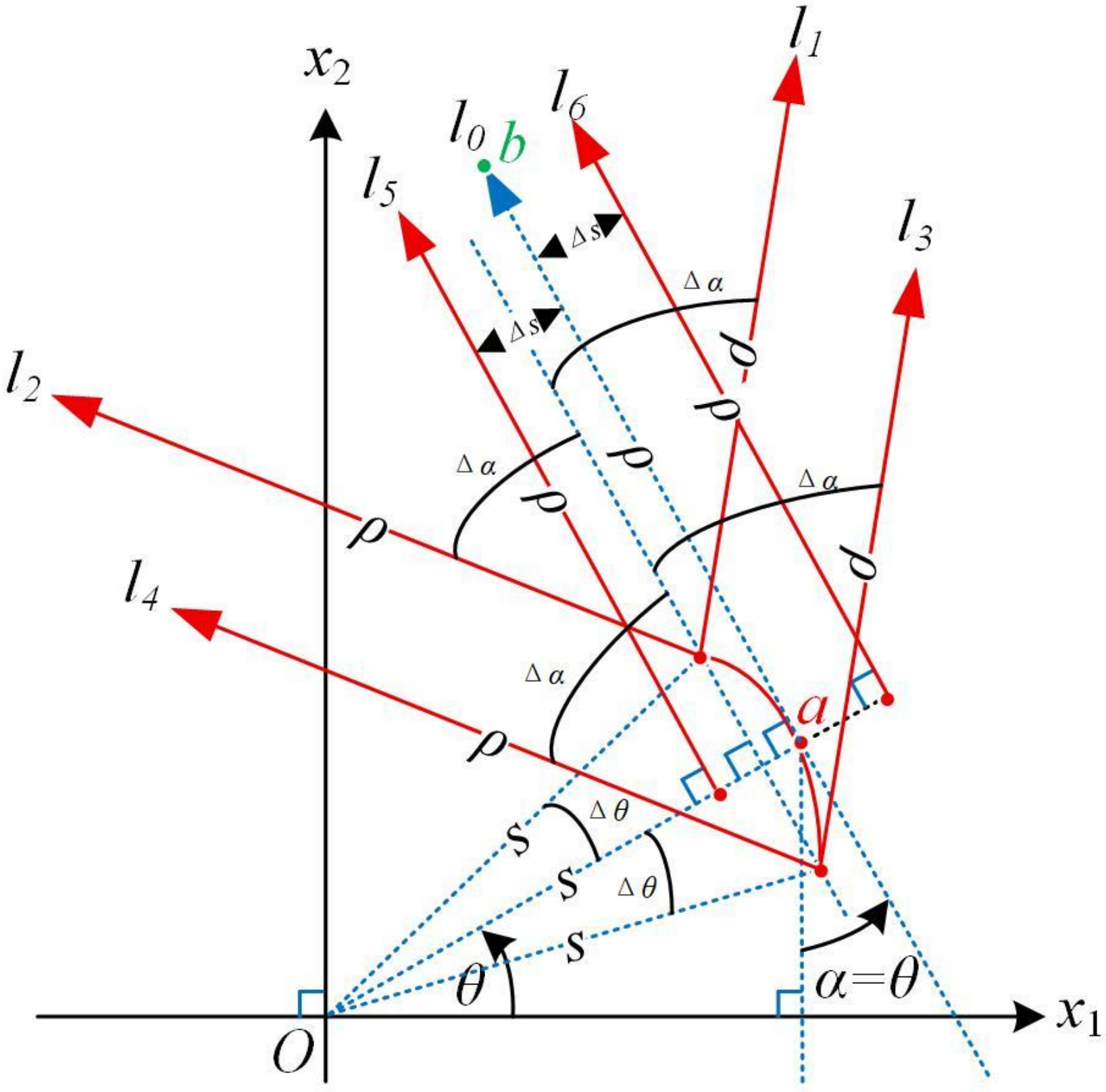

Fig. 2 Physical meaning of formula (6.6) is $((l_2\text{-}l_1)\text{-}(l_4\text{-}l_3))/(\Delta\alpha\Delta\theta) \approx 2s(l_6\text{-}l_5)/\Delta s$, wherein $l_i$ represents the line integral corresponding to the *ith* X-ray path.

### 7.2 Formula (6.7): $\partial^2 G/\partial s\partial\beta = -\partial^2 G/\partial z_0\partial\alpha$

This formula establishes an equality relation between two different mixed partial derivatives. Left-hand side: $\partial^2 G/\partial s\partial\beta$ represents the mixed rate of change of the projection value with respect to the source point radial distance $s$ and the ray tilt angle $\beta$. This describes the change in the projection value when both the source point radial position and the ray tilt angle vary simultaneously. Right-hand side: $-\partial^2 G/\partial z_0\partial\alpha$ represents the mixed rate of change of the projection value with respect to the source point axial position $z_0$ and the ray direction azimuthal angle $\alpha$.

Physical meaning: The effects on the projection data of the joint variation of source point radial-tilt and the joint variation of source point axial-ray azimuth are equivalent. This symmetry reflects the geometric symmetry of CT systems: variations in the horizontal plane ($s$ and $\beta$) and variations in the vertical direction ($z_0$ and $\alpha$) are profoundly connected.

### 7.3 Formula (6.8): $\partial^2 G/\partial\theta\partial\beta = -s\partial G/\partial z_0$

This formula relates the source point azimuthal angle $\theta$, the ray tilt angle $\beta$, and the source point axial position $z_0$. Left-hand side: $\partial^2 G/\partial\theta\partial\beta$ represents the mixed rate of change of the projection value with respect to the source point azimuthal angle and the ray tilt angle. Right-hand side: $-s\partial G/\partial z_0$ represents the rate of change of the projection value with respect to the source point axial position, multiplied by $-s$.

Physical meaning: The variation characteristics of the projection data in the azimuth-tilt joint direction are directly related to its variation characteristics in the axial direction. In helical CT, the source point simultaneously undergoes rotation ($\theta$ variation) and translation ($z_0$ variation); this formula describes the coupled effects of these two motions on the projection data. For cone-beam CT, this equation is particularly meaningful because it relates the cone-angle effect (through $\beta$) to the axial translation sampling effect (through $z_0$). It is particularly noteworthy that this equation strictly holds only when $\alpha = \theta$ and $\beta = 0$.

## 8. Discussion

The specific forms of the John equation derived in this paper have broad application prospects in the field of CT imaging.

### 8.1 Data Consistency Verification

In practical CT systems, measured projection data inevitably contains inconsistencies such as noise, scattering [23], beam-hardening artifacts [20-22], and motion artifacts [19]. The John equation provides a strict data consistency criterion: any physically reasonable projection data must satisfy these equations. Therefore, one can assess data quality by checking whether the measured data satisfies (or approximately satisfies) the John equation, identify inconsistent data portions, and perform correction.

In specific implementation, the corresponding partial derivatives of the projection data can be computed numerically, and the degree to which the John equation holds can be checked. The degree of deviation can be quantified as a data inconsistency index to guide data correction or the selection of reconstruction algorithms.

### 8.2 Geometric Parameter Calibration

Precise calibration of geometric parameters (such as source-axis distance, ray direction parameters, rotation center, etc.) in CT systems is crucial for high-quality reconstruction [26]. However, in practical systems, these parameters may have errors. The John equation provides the possibility of calibrating geometric parameters using the projection data itself.

The basic idea is: if the geometric parameters are accurate, the projection data should accurately satisfy the John equation; if the parameters are incorrect, the equation does not hold. Therefore, geometric parameters can be optimized so that the projection data satisfies the John equation as much as possible, thereby achieving self-calibration of the parameters. This method does not rely on external phantoms and uses only the projection data of the scanned object, having important practical value.

### 8.3 Incomplete-Data Reconstruction

In practical applications, incomplete-data problems are often encountered, such as limited-angle scanning, sparse-angle sampling, or truncated data. Traditional reconstruction algorithms may fail or produce severe artifacts under these conditions. The John equation provides a new approach to constrain incomplete-data reconstruction. This constrained reconstruction method uses the internal consistency of the data itself to compensate for missing information, potentially improving reconstruction quality under incomplete-data conditions [17,18].

### 8.4 Novel Scanning Trajectory Design

The John equation reveals the intrinsic relationships between projection data under variations of different geometric parameters, providing theoretical guidance for designing novel scanning trajectories. For example, equations (6.7) and (6.8) show that certain complex scans may be approximately equivalent to simpler scans, thereby simplifying system design or reducing scanning time.

### 8.5 Theoretical and Computational Challenges

Although this paper derives the specific form of the John equation under mixed parameterization, some challenges remain in practical applications:

*(i)* Numerical stability: Partial derivative computation is sensitive to noise, requiring robust numerical differentiation methods;

*(ii)* Computational complexity: The computational cost of calculating high-order partial derivatives and high-dimensional optimization problems is high;

*(iii)* Model error: Actual physical effects (such as beam hardening, scattering, noise, non-monochromatic spectrum, etc.) may cause projection data to not completely satisfy the ideal John equation;

*(iv)* Boundary effects: At data boundaries, the computation of partial derivatives requires special treatment.

Future research needs to develop effective numerical algorithms and correction models for these challenges.

### 8.6 Relationship with Related Work

Under the special geometry of $\alpha = \theta$, the complete constraints satisfied by the projection data $G$ consist of a first-order PDE (translation invariance condition (D.3)) and a second-order PDE (such as (6.4)). The remaining formulas (such as (6.1), (6.5)) are all corollaries thereof. This structure reveals that the John equation provides second-order coupling in different parameter directions, while translation invariance gives parameter redundancy in the ray direction (first-order); both are indispensable.

**Table 1 Constraint Structures of the Original John Equation, Patch's Result and this work**

| Theoretical Framework | Independent Equations | Number of Variables | Description |
|---|---|---|---|
| Original John Equation [3] | 3 second-order PDEs | 6 (3 components each for $a$, $b$) | 3D Cartesian coordinates, no parameterization reduction |
| This work ($\alpha = \theta$ cone-beam geometry) | 1 first-order PDE (translation invariance), 1 second-order PDE | 5 ($\rho$ eliminated) | -1 homogeneity eliminates $\rho$, and translation invariance condition provides a first-order constraint |
| Patch's Result [9] (third-generation cone-beam geometry) | 1 second-order PDE | 4 ($\rho$, $d$ eliminated in [9]) | -1 homogeneity used to eliminate two parameters |

The original John equation has 6 independent variables and 3 independent second-order PDEs [3]; the result of this paper has 5 independent variables ($\rho$ can be eliminated), 1 -1 homogeneity condition (first-order PDE), 1 translation invariance condition (first-order PDE), and 1 independent second-order PDE; Patch's result has 4 independent variables (another 2 can be eliminated), 2 -1 homogeneity conditions (first-order PDEs), and 1 independent second-order PDE [9]. See the independence summary in Appendix E and the comparison with classical results. As shown in Table 1, as the parameterization gradually approaches the actual scanning geometry, the order and number of independent constraints show a trend of reduction and simplification, which is the inevitable result of geometric redundancy being progressively explicitly separated.

## 9. Conclusion

This paper systematically studied the John equation constraints for the 3D X-ray transform under mixed parameterization. Through an innovative parameterization scheme—the source point using cylindrical coordinates $a = (s \cos \theta, s \sin \theta, z_0)$ and the ray direction using spherical coordinates $d = \rho(-\cos \beta \sin \alpha, \cos \beta \cos \alpha, \sin \beta)$—a bridge between the abstract John equation and concrete CT geometry was established. The main contributions are summarized as follows:

*(i)* Complete mathematical derivation: The transformation formulas for partial derivative operators from Cartesian coordinates to mixed parameters were established, transforming the ray-direction-based John equation into a specific form in the new parameter system, and using -1 homogeneity to eliminate partial derivatives with respect to $\rho$.

*(ii)* Simplification under special geometric configurations: Under the common CT configuration of $\alpha = \theta$ and $\beta = 0$, three concise differential constraint equations were obtained.

*(iii)* In-depth physical interpretation: The geometric meaning and physical connotation of each equation were analyzed in detail, revealing the intrinsic correlations of CT projection data under variations of different parameters.

*(iv)* Broad application prospects: The potential applications of the John equation in data consistency verification, geometric parameter calibration, incomplete-data reconstruction, and scanning trajectory design were discussed.

The research results of this paper not only deepen the understanding of the mathematical theory of the X-ray transform, but also provide new ideas for solving practical problems in CT imaging. Future work will focus on the development of numerical algorithms, validation with actual data, and the extension of the equations under more complex geometric configurations.

## Acknowledgments

We would like to express our sincere gratitude to Dr. Ge Wang for his carefully checking the manuscript; to Dr. Tianye Niu and Dr. Ming Chen for the helpful discussion on the manuscript; and to Senyang Xiong, Yu Zhang and Yijun Qiao for their assistances in editing the manuscript.

# Appendices

## Appendix A: Derivation of the John Equation Based on Ray Direction

From the definition $d = b - a$, we get $b = a + d$, and $g(a, d) = f(a, a + d) = f(a, b)$. When $b$ is fixed, $d = b - a$ varies with $a$. Applying the chain rule to the composite function $g(a, d(a))$:

$$\partial/\partial a_i|_b = \partial/\partial a_i|_d + \Sigma_{k=1}^3 (\partial/\partial d_k) \cdot (\partial d_k/\partial a_i) \tag{A.1}$$

From $d_k = b_k - a_k$ and $b$ fixed, we get $\frac{\partial d_k}{\partial a_i} = -\delta_{ik}$. Substituting into the above equation:

$$\partial/\partial a_i|_b = \partial/\partial a_i|_d - \partial/\partial d_i \tag{A.2}$$

From $f(a, b) = g(a, d)$ we get

$$\partial f/\partial b_j = \partial g/\partial d_j, \quad \partial f/\partial a_i|_b = \partial g/\partial a_i|_d - \partial g/\partial d_i \tag{A.3}$$

Substituting into the left-hand side of (2.2):

$$\partial/\partial a_i|_b\ (\partial f/\partial b_j) = (\partial/\partial a_i|_d - \partial/\partial d_i)\ \partial g/\partial d_j = \partial^2 g/\partial a_i\partial d_j - \partial^2 g/\partial d_i\partial d_j \tag{A.4}$$

Substituting into the right-hand side of (2.2):

$$\partial/\partial a_j|_b\ (\partial f/\partial b_i) = (\partial/\partial a_j|_d - \partial/\partial d_j)\ \partial g/\partial d_i = \partial^2 g/\partial a_j\partial d_i - \partial^2 g/\partial d_j\partial d_i \tag{A.5}$$

Since $\partial^2 g/\partial d_i \partial d_j = \partial^2 g/\partial d_j \partial d_i$, subtracting the two equations yields:

$$\partial^2 g/\partial a_i \partial d_j = \partial^2 g/\partial a_j \partial d_i \quad \text{(A.6)}$$

## Appendix B: Partial Derivative Transformation with Respect to Source Point Coordinates

By the chain rule, the partial derivative with respect to the source point coordinate $a_i$ can be expressed as:

$$\partial/\partial a_i = (\partial s/\partial a_i)(\partial/\partial s) + (\partial\theta/\partial a_i)(\partial/\partial\theta) + (\partial z_0/\partial a_i)(\partial/\partial z_0) \quad \text{(B.1)}$$

According to the cylindrical coordinate relations (3.1), each partial derivative can be calculated. For $a_1$:

$$\partial s/\partial a_1 = a_1/s = \cos\theta; \quad \partial\theta/\partial a_1 = -a_2/s^2 = -(\sin\theta)/s; \quad \partial z_0/\partial a_1 = 0 \quad \text{(B.2)}$$

Therefore:

$$\partial/\partial a_1 = \cos\theta\, \partial/\partial s - (\sin\theta/s)\, \partial/\partial\theta \quad \text{(B.3)}$$

Similarly, for $a_2$:

$$\partial s/\partial a_2 = a_2/s = \sin\theta; \quad \partial\theta/\partial a_2 = a_1/s^2 = \cos\theta/s; \quad \partial z_0/\partial a_2 = 0 \quad \text{(B.4)}$$

Therefore:

$$\partial/\partial a_2 = \sin\theta\, \partial/\partial s + (\cos\theta/s)\, \partial/\partial\theta \quad \text{(B.5)}$$

For $a_3$:

$$\partial s/\partial a_3 = 0; \quad \partial\theta/\partial a_3 = 0; \quad \partial z_0/\partial a_3 = 1 \quad \text{(B.6)}$$

Therefore:

$$\partial/\partial a_3 = \partial/\partial z_0 \quad \text{(B.7)}$$

## Appendix C: Partial Derivative Transformation with Respect to Ray Direction Coordinates

The partial derivative transformation with respect to the ray direction coordinate $d_j$ requires computing the inverse Jacobian matrix from $(d_1, d_2, d_3)$ to $(\rho, \alpha, \beta)$. First compute the Jacobian matrix $J$, whose elements are $\partial d_i/\partial\eta_j$, where $\eta = (\rho, \alpha, \beta)$:

$$J = [\ \partial d_1/\partial\rho \ \ \partial d_1/\partial\alpha \ \ \partial d_1/\partial\beta$$

$$\partial d_2/\partial\rho \ \ \partial d_2/\partial\alpha \ \ \partial d_2/\partial\beta$$

$$\partial d_3/\partial\rho \ \ \partial d_3/\partial\alpha \ \ \partial d_3/\partial\beta\ ] \quad \text{(C.1)}$$

From equation (3.2), compute each partial derivative:

$$\partial d_1/\partial\rho = -\cos\beta \sin\alpha; \quad \partial d_1/\partial\alpha = -\rho \cos\beta \cos\alpha; \quad \partial d_1/\partial\beta = \rho \sin\beta \sin\alpha;$$

$$\partial d_2/\partial\rho = \cos\beta \cos\alpha; \quad \partial d_2/\partial\alpha = -\rho \cos\beta \sin\alpha; \quad \partial d_2/\partial\beta = -\rho \sin\beta \cos\alpha;$$

$$\partial d_3/\partial\rho = \sin\beta; \quad \partial d_3/\partial\alpha = 0; \quad \partial d_3/\partial\beta = \rho \cos\beta \tag{C.2}$$

Therefore:

$$J = [\, -\cos\beta \sin\alpha \quad -\rho \cos\beta \cos\alpha \quad \rho \sin\beta \sin\alpha$$

$$\cos\beta \cos\alpha \quad -\rho \cos\beta \sin\alpha \quad -\rho \sin\beta \cos\alpha$$

$$\sin\beta \quad 0 \quad \rho \cos\beta \,] \tag{C.3}$$

The Jacobian determinant is:

$$det(J) = \rho^2 \cos\beta \tag{C.4}$$

Since $\rho > 0$ and $\beta \in (-\pi/2, \pi/2)$, we have cos $\beta > 0$, so the Jacobian matrix is invertible.

By computing the inverse matrix or directly using geometric relations, we obtain:

$$\partial\rho/\partial d_1 = d_1/\rho = -\cos\beta \sin\alpha; \; \partial\rho/\partial d_2 = d_2/\rho = \cos\beta \cos\alpha; \; \partial\rho/\partial d_3 = d_3/\rho = \sin\beta \tag{C.5}$$

For $\alpha$, from the relation $tan\ \alpha = -d_1/d_2$ we get:

$$\partial\alpha/\partial d_1 = -d_2/(d_1^2 + d_2^2) = -\cos\alpha/(\rho \cos\beta);$$

$$\partial\alpha/\partial d_2 = d_1/(d_1^2 + d_2^2) = -\sin\alpha/(\rho \cos\beta);$$

$$\partial\alpha/\partial d_3 = 0 \tag{C.6}$$

For $\beta$, from the relation $sin\ \beta = d_3/\rho$ we get:

$$\partial\beta/\partial d_1 = -d_1 d_3/(\rho^3 \cos\beta) = \sin\beta \sin\alpha/\rho;$$

$$\partial\beta/\partial d_2 = -d_2 d_3/(\rho^3 \cos\beta) = -\sin\beta \cos\alpha/\rho;$$

$$\partial\beta/\partial d_3 = \cos\beta/\rho \tag{C.7}$$

By the chain rule, the partial derivative with respect to the ray direction coordinate $d_i$ can be expressed as:

$$\partial/\partial d_i = (\partial\rho/\partial d_i)(\partial/\partial\rho) + (\partial\alpha/\partial d_i)(\partial/\partial\alpha) + (\partial\beta/\partial d_i)(\partial/\partial\beta) \tag{C.8}$$

Each partial derivative can be computed:

$$\partial/\partial d_1 = -\cos\beta \sin\alpha\, \partial/\partial\rho - \cos\alpha/(\rho\cos\beta)\, \partial/\partial\alpha + \sin\beta \sin\alpha/\rho\, \partial/\partial\beta \tag{C.9}$$

$$\partial/\partial d_2 = \cos\beta \cos\alpha\, \partial/\partial\rho - \sin\alpha/(\rho\cos\beta)\, \partial/\partial\alpha - \sin\beta \cos\alpha/\rho\, \partial/\partial\beta \tag{C.10}$$

$$\partial/\partial d_3 = \sin\beta\, \partial/\partial\rho + (\cos\beta/\rho)\, \partial/\partial\beta \tag{C.11}$$

## Appendix D: Relationship Between Translation Invariance Condition and John Equation

**Definition: Translation Invariance Condition**

The X-ray transform has a fundamental geometric property: if the source point $a$ is moved an arbitrary distance along the ray direction $d$, the projection value remains unchanged. That is, for any real number $\tau$, we have

$$f(a + \tau d, b + \tau d) = f(a, b) \tag{D.1}$$

Under the $(a, d)$ parameterization, letting $g(a, d) = f(a, a + d)$, then (D.1) is equivalent to $g(a + \tau d, d) = g(a, d)$. Differentiating with respect to $\tau$ and setting $\tau = 0$, we obtain the first-order linear PDE:

$$\Sigma_{i=1}^{3} d_i\, \partial g/\partial a_i = 0 \tag{D.2}$$

This is the differential form of the translation invariance condition. Using formulas (3.2), (4.2)-(4.4), and eliminating $\rho$, we get:

$$\cos\beta \sin(\theta - \alpha)\, \partial G/\partial s + (\cos\beta/s) \cos(\theta - \alpha)\, \partial G/\partial\theta + \sin\beta\, \partial G/\partial z_0 = 0 \tag{D.3}$$

Under $\alpha = \theta$, this simplifies to:

$$(\cos\beta/s)\, \partial G/\partial\theta + \sin\beta\, \partial G/\partial z_0 = 0 \tag{D.4}$$

Differentiating formula (D.3) with respect to $\alpha$ yields:

$$-\cos\beta \cos(\theta - \alpha)\, \partial G/\partial s + \cos\beta \sin(\theta - \alpha)\, \partial^2 G/\partial s\partial\alpha + (\cos\beta/s) \sin(\theta - \alpha)\, \partial G/\partial\theta$$

$$+ (\cos\beta/s) \cos(\theta - \alpha)\, \partial^2 G/\partial\theta\partial\alpha + \sin\beta\, \partial^2 G/\partial z_0\partial\alpha = 0 \tag{D.5}$$

Under $\alpha = \theta$, this simplifies to:

$$-s\, \partial G/\partial s + \partial^2 G/\partial\theta\partial\alpha + s \tan\beta\, \partial^2 G/\partial z_0\partial\alpha = 0 \tag{D.6}$$

Below we analyze the relationship between the translation invariance condition and the formulas in the main text; note that all derivations are ultimately discussed under the condition $\alpha = \theta$.

**Proposition D.1: The translation invariance condition (D.4) can derive formula (6.5)**

Proof: Differentiate formula (D.4) with respect to $\beta$:

$$-(\sin\beta/s)\ \partial G/\partial\theta + (\cos\beta/s)\ \partial^2 G/\partial\theta\partial\beta + \cos\beta\ \partial G/\partial z_0 + \sin\beta\ \partial^2 G/\partial z_0\partial\beta = 0$$

Multiplying by $s/\cos\beta$ yields formula (6.5).

**Proposition D.2: The translation invariance condition (D.6) combined with formula (6.1) can derive formula (6.4)**

Proof: Substitute formula (6.1) into the translation invariance condition formula (D.6), eliminating the $\partial^2 G/\partial\theta\partial\alpha$ term to get:

$$-s\sin^2\beta\ \partial G/\partial s + s\cos\beta\sin\beta\ \partial^2 G/\partial s\partial\beta + s\tan\beta\ \partial^2 G/\partial z_0\partial\alpha = 0$$

Multiplying by $\cot\beta/s$ yields formula (6.4).

**Proposition D.3: The translation invariance condition (D.6) combined with formula (6.4) can derive formula (6.1)**

Proof: Substitute formula (6.4) into the translation invariance condition formula (D.6), eliminating the $\partial^2 G/\partial z_0\partial\alpha$ term to get:

$$(\cos\beta/s)\ \partial^2 G/\partial\theta\partial\alpha = \cos\beta\ \partial G/\partial s - \sin\beta\ (\cos\beta\sin\beta\ \partial G/\partial s - \cos^2\beta\ \partial^2 G/\partial s\partial\beta)$$

Multiplying both sides by $s/\cos\beta$ yields formula (6.1).

**Proposition D.4: Formula (6.1) combined with (6.4) can derive the translation invariance condition (D.6)**

Proof: From formula (6.4) we get

$$s\cos\beta\sin\beta\ \partial^2 G/\partial s\partial\beta = s\sin^2\beta\ \partial G/\partial s - s\tan\beta\ \partial^2 G/\partial z_0\partial\alpha$$

Substituting into formula (6.1) yields the translation invariance condition (D.6).

**Proposition D.5: Formula (6.1) combined with (5.13) can derive formula (6.5)**

Proof: Differentiate both sides of formula (5.13) with respect to $\theta$, and set $\alpha = \theta$ to get:

$$-\tan\beta\ \partial^2 G/\partial s\partial\theta + \partial^3 G/\partial s\partial\beta\partial\theta = -\partial G/\partial z_0 - (1/\cos^2\beta)\ \partial^3 G/\partial z_0\partial\alpha\partial\theta - \tan\beta\ \partial^2 G/\partial z_0\partial\beta \quad \text{(D.7)}$$

Note that this differs from formula (6.4). Then differentiate both sides of formula (6.1) with respect to $z_0$ to get:

$$(1/\cos^2\beta)\ \partial^3 G/\partial\theta\partial\alpha\partial z_0 = s\ \partial^2 G/\partial s\partial z_0 + s\tan\beta\ \partial^3 G/\partial s\partial\beta\partial z_0$$

After substituting into the right-hand side of (D.7) and rearranging:

$$-\tan\beta\ \partial^2 G/\partial s\partial\theta + \partial^3 G/\partial s\partial\beta\partial\theta = (-\partial G/\partial z_0 - s\ \partial^2 G/\partial s\partial z_0)$$

$$+ \tan\beta\ (-s\ \partial^3 G/\partial s\partial\beta\partial z_0 - \partial^2 G/\partial z_0\partial\beta) \qquad \text{(D.8)}$$

Considering that the function $G$ satisfies compact support boundary conditions (all partial derivatives are zero when $s \to -\infty$), integrate both sides of (D.7) with respect to $s$ from $-\infty$ to $s$. Integration of the first term on the left-hand side:

$$\int_{-\infty}^{s} -\tan\beta\ \partial^2 G/\partial s'\partial\theta\ ds' = -\tan\beta\ \partial G/\partial\theta$$

Integration of the second term on the left-hand side:

$$\int_{-\infty}^{s} \partial^3 G/\partial s'\partial\beta\partial\theta\ ds' = \partial^2 G/\partial\beta\partial\theta$$

Integration of the first term on the right-hand side (integration by parts):

$$\int_{-\infty}^{s} (-\partial G/\partial z_0 - s'\ \partial^2 G/\partial s'\partial z_0)\ ds' = -s\ \partial G/\partial z_0$$

Integration of the second term on the right-hand side (integration by parts):

$$\tan\beta \int_{-\infty}^{s} (-s'\ \partial^3 G/\partial s'\partial\beta\partial z_0 - \partial^2 G/\partial z_0\partial\beta)\ ds' = -s\ \tan\beta\ \partial^2 G/\partial\beta\partial z_0$$

Finally obtaining the equation without integral terms:

$$-\tan\beta\ \partial G/\partial\theta + \partial^2 G/\partial\beta\partial\theta = -s\ \partial G/\partial z_0 - s\ \tan\beta\ \partial^2 G/\partial\beta\partial z_0$$

which is formula (6.5).

## Appendix E: MATLAB Code for Validating Formula (D.3)

```
%% Verify Eq. (D.3): cosβ sin(θ-α) ∂G/∂s + (cosβ/s) cos(θ-α) ∂G/∂θ + sinβ ∂G/∂z0 = 0

% Compute line integrals using the Shepp-Logan phantom

close all; clear all; clc;


%% 1. Parameter setup

s0 = 0.5;         % radial distance (must be non-zero)

z0 = 0.1;         % axial position

ds = 0.01;        % step size for ∂/∂s

dz0 = 0.01;       % step size for ∂/∂z0

rho = 1;          % ray length (fixed to 1 by -1 homogeneity)
```

```
theta = linspace(-pi/4, pi/4, 21);   % source azimuth angle
alpha = linspace(-pi/4, pi/4, 21);   % ray azimuth angle (independent of θ)
beta  = linspace(-pi/4, pi/4, 21);   % ray tilt angle

dtheta = theta(2) - theta(1);
dalpha = alpha(2) - alpha(1);   % currently α differencing is not needed
dbeta  = beta(2) - beta(1);

n_theta = length(theta);
n_alpha = length(alpha);
n_beta  = length(beta);

fprintf('Validating Eq. (D.3): cosβ sin(θ-α) ∂G/∂s + (cosβ/s) cos(θ-α) ∂G/∂θ + sinβ ∂G/∂z0 = 0\n');
fprintf('s0 = %.3f, ds = %.4f, z0 = %.3f, dz0 = %.4f\n', s0, ds, z0, dz0);
fprintf('Number of θ points: %d, α points: %d, β points: %d\n', n_theta, n_alpha, n_beta);

%% 2. Preallocate storage arrays
% G0:         G(s0, θ, z0, α, β)
% G_sp:       G(s0+ds, θ, z0, α, β)
% G_sm:       G(s0-ds, θ, z0, α, β)
% G_tp:       G(s0, θ+dθ, z0, α, β)
% G_tm:       G(s0, θ-dθ, z0, α, β)
% G_zp:       G(s0, θ, z0+dz0, α, β)
% G_zm:       G(s0, θ, z0-dz0, α, β)

G0   = zeros(n_beta, n_alpha, n_theta);
G_sp = zeros(n_beta, n_alpha, n_theta);
```

```
G_sm = zeros(n_beta, n_alpha, n_theta);
G_tp = zeros(n_beta, n_alpha, n_theta);
G_tm = zeros(n_beta, n_alpha, n_theta);
G_zp = zeros(n_beta, n_alpha, n_theta);
G_zm = zeros(n_beta, n_alpha, n_theta);

fprintf('Calculating required G values...\n');
for ib = 1:n_beta
    b = beta(ib);
    for ia = 1:n_alpha
        a = alpha(ia);
        for it = 1:n_theta
            t = theta(it);

            % Base source position (s0, t, z0)
            x0 = s0 * cos(t);
            y0 = s0 * sin(t);

            % Ray direction (fixed α, β, ρ)
            dx = -cos(b) * sin(a);
            dy =  cos(b) * cos(a);
            dz =  sin(b);

            % Center point
            G0(ib,ia,it) = line_integral_from_source(x0, y0, z0, dx, dy, dz);

            % s+ds
```

```
    x_sp = (s0+ds) * cos(t);
    y_sp = (s0+ds) * sin(t);
    G_sp(ib,ia,it) = line_integral_from_source(x_sp, y_sp, z0, dx, dy, dz);

    % s-ds
    x_sm = (s0-ds) * cos(t);
    y_sm = (s0-ds) * sin(t);
    G_sm(ib,ia,it) = line_integral_from_source(x_sm, y_sm, z0, dx, dy, dz);

    % θ+dθ (source changes, direction stays the same)
    t_p = t + dtheta;
    x_tp = s0 * cos(t_p);
    y_tp = s0 * sin(t_p);
    G_tp(ib,ia,it) = line_integral_from_source(x_tp, y_tp, z0, dx, dy, dz);

    % θ-dθ
    t_m = t - dtheta;
    x_tm = s0 * cos(t_m);
    y_tm = s0 * sin(t_m);
    G_tm(ib,ia,it) = line_integral_from_source(x_tm, y_tm, z0, dx, dy, dz);

    % z0+dz0
    G_zp(ib,ia,it) = line_integral_from_source(x0, y0, z0+dz0, dx, dy, dz);

    % z0-dz0
    G_zm(ib,ia,it) = line_integral_from_source(x0, y0, z0-dz0, dx, dy, dz);
end
```

```
    end
end

%% 3. Compute partial derivatives (central differences)
% ∂G/∂s
dG_ds = (G_sp - G_sm) / (2*ds);

% ∂G/∂θ
dG_dtheta = (G_tp - G_tm) / (2*dtheta);

% ∂G/∂z0
dG_dz0 = (G_zp - G_zm) / (2*dz0);

%% 4. Verify whether the left-hand side of Eq. (D.3) equals zero
fprintf('\nVerification results (only showing non-zero data points with G0 > 1e-6):\n');
fprintf('------------------------------------------------------------------------------------------------------------\n');
fprintf('   θ        α        β          G0            LHS       RHS    error\n');
fprintf('------------------------------------------------------------------------------------------------------------\n');

total_err = 0;
count = 0;
max_err = 0;

for ib = 1:n_beta
    b = beta(ib);
    for ia = 1:n_alpha
        a = alpha(ia);
```

```
        for it = 1:n_theta
            t = theta(it);
            g0 = G0(ib,ia,it);
            if g0 < 1e-6   % skip points where the ray barely passes through the object
                continue;
            end

            % Compute LHS: cosβ sin(θ-α) ∂G/∂s + (cosβ/s) cos(θ-α) ∂G/∂θ + sinβ ∂G/∂z0
            LHS = cos(b)*sin(t-a) * dG_ds(ib,ia,it) ...
                + (cos(b)/s0)*cos(t-a) * dG_dtheta(ib,ia,it);
            % Compute RHS: - sinβ ∂G/∂z0
            RHS = - sin(b) * dG_dz0(ib,ia,it);

            err = abs(LHS - RHS);
            fprintf('%8.3f %8.3f %8.3f %12.4e %12.4e %12.4e %12.4e\n', t, a, b, g0, LHS,RHS,err);

            total_err = total_err + err;
            if err > max_err, max_err = err; end
            count = count + 1;
        end
    end
end

fprintf('------------------------------------------------------------------------------------------------------------\n');
if count > 0
    fprintf('Number of valid points: %d\n', count);
    fprintf('Mean absolute error: %e\n', total_err/count);
```

```
    fprintf('Maximum absolute error: %e\n', max_err);
else
    fprintf('Warning: No valid points where the ray passes through the object. Please adjust parameters or check the phantom definition.\n');
end


%% Helper function: line integral through the Shepp-Logan phantom (from a source point, direction is normalized)
function u = line_integral_from_source(x0, y0, z0, dx, dy, dz)
    % Shepp-Logan phantom parameters [center_x, center_y, center_z, semi-axis a, semi-axis b, semi-axis c, density]
    ellipsoids = [
        0.3,   0.1,   0,     0.69,   0.92,   0.9,    2.0;
        0,     0,     0,     0.6624, 0.874,  0.88,  -0.98;
       -0.22,  0,    -0.25,  0.41,   0.16,   0.21,  -0.2;
        0.22,  0,    -0.25,  0.31,   0.11,   0.22,  -0.2;
        0,     0.35, -0.25,  0.21,   0.25,   0.5,    0.2;
        0,     0.1,   0.625, 0.046,  0.046,  0.046,  0.2;
       -0.08, -0.605,0,      0.046,  0.023,  0.02,   0.1;
        0.06, -0.605,0,      0.046,  0.023,  0.02,   0.1;
        0,    -0.1,   0.625, 0.056,  0.04,   0.1,    0.2;
        0.06, -0.105, 0.625, 0.056,  0.056,  0.1,   -0.2;
    ];


    u = 0;
    tol = 1e-12;
    for k = 1:size(ellipsoids,1)
        xc = ellipsoids(k,1); yc = ellipsoids(k,2); zc = ellipsoids(k,3);
```

```
        a  = ellipsoids(k,4); b  = ellipsoids(k,5); c  = ellipsoids(k,6);
        density = ellipsoids(k,7);

        A = dx^2/a^2 + dy^2/b^2 + dz^2/c^2;
        if abs(A) < tol
            continue;
        end
        B = 2*( dx*(x0-xc)/a^2 + dy*(y0-yc)/b^2 + dz*(z0-zc)/c^2 );
        C = (x0-xc)^2/a^2 + (y0-yc)^2/b^2 + (z0-zc)^2/c^2 - 1;

        disc = B^2 - 4*A*C;
        if disc > tol
            sqrt_disc = sqrt(disc);
            t1 = (-B - sqrt_disc) / (2*A);
            t2 = (-B + sqrt_disc) / (2*A);
            u = u + density * (t2 - t1);
        end
    end
end
```

## Appendix F: MATLAB Code for Validating Formula (6.1)

```
%% Verify Eq. (6.1): ∂²G/∂θ∂α = s cos²β ∂G/∂s + s cosβ sinβ ∂²G/∂s∂β   (α=θ)
% Compute line integrals through the Shepp-Logan phantom

close all; clear all; clc;

%% 1. Parameter settings
s0 = 0.5;               % source radial distance
```

```
ds = 0.01;              % step size for ∂/∂s
z0 = 0.1;               % source axial position
rho = 1;                % ray length (fixed by -1 homogeneity)

theta = linspace(-pi/4, pi/4, 41);  % source azimuth (also baseline α=θ)
beta  = linspace(-pi/4, pi/4, 41);  % ray tilt angle

dtheta = theta(2) - theta(1);
dbeta  = beta(2)  - beta(1);
dalpha = dtheta;        % α step same as θ step

n_theta = length(theta);
n_beta  = length(beta);

fprintf('Verify formula (6.1): ∂²G/∂θ∂α = s cos²β ∂G/∂s + s cosβ sinβ ∂²G/∂s∂β\n');
fprintf('s0 = %.3f, ds = %.3f, z0 = %.2f\n', s0, ds, z0);
fprintf('Number of θ points: %d, Number of β points: %d\n', n_theta, n_beta);

%% 2. Compute required G values
% For each (beta, theta) compute 11 auxiliary G values:
% G0         : (s0, θ, α=θ, β)
% G_tp_ap    : (s0, θ+dθ, α=θ+dα, β)
% G_tp_am    : (s0, θ+dθ, α=θ-dα, β)
% G_tm_ap    : (s0, θ-dθ, α=θ+dα, β)
% G_tm_am    : (s0, θ-dθ, α=θ-dα, β)
% G_sp       : (s0+ds, θ, α=θ, β)
% G_sm       : (s0-ds, θ, α=θ, β)
```

```
% G_sp_bp    : (s0+ds, θ, α=θ, β+dbeta)
% G_sp_bm    : (s0+ds, θ, α=θ, β-dbeta)
% G_sm_bp    : (s0-ds, θ, α=θ, β+dbeta)
% G_sm_bm    : (s0-ds, θ, α=θ, β-dbeta)

% Preallocation
G0      = zeros(n_beta, n_theta);
G_tp_ap = zeros(n_beta, n_theta);
G_tp_am = zeros(n_beta, n_theta);
G_tm_ap = zeros(n_beta, n_theta);
G_tm_am = zeros(n_beta, n_theta);
G_sp    = zeros(n_beta, n_theta);
G_sm    = zeros(n_beta, n_theta);
G_sp_bp = zeros(n_beta, n_theta);
G_sp_bm = zeros(n_beta, n_theta);
G_sm_bp = zeros(n_beta, n_theta);
G_sm_bm = zeros(n_beta, n_theta);

fprintf('Computing required G values...\n');
for i = 1:n_beta
    b = beta(i);
    for j = 1:n_theta
        t = theta(j);

        % ---- Baseline source and direction (α=θ) ----
        x0 = s0 * cos(t);
        y0 = s0 * sin(t);
```

```
dx = -cos(b) * sin(t);
dy =  cos(b) * cos(t);
dz =  sin(b);

% G0
G0(i,j) = line_integral_from_source(x0, y0, z0, dx, dy, dz);

% ---- θ+dθ, α=θ+dα ----
t_p = t + dtheta;
a_p = t + dalpha;    % α=θ+dα
x_tp = s0 * cos(t_p);
y_tp = s0 * sin(t_p);
dx_tp_ap = -cos(b) * sin(a_p);
dy_tp_ap =  cos(b) * cos(a_p);
dz_tp_ap =  sin(b);
G_tp_ap(i,j) = line_integral_from_source(x_tp, y_tp, z0, dx_tp_ap, dy_tp_ap, dz_tp_ap);

% ---- θ+dθ, α=θ-dα ----
a_m = t - dalpha;
dx_tp_am = -cos(b) * sin(a_m);
dy_tp_am =  cos(b) * cos(a_m);
dz_tp_am =  sin(b);
G_tp_am(i,j) = line_integral_from_source(x_tp, y_tp, z0, dx_tp_am, dy_tp_am, dz_tp_am);

% ---- θ-dθ, α=θ+dα ----
t_m = t - dtheta;
x_tm = s0 * cos(t_m);
```

```
y_tm = s0 * sin(t_m);

dx_tm_ap = -cos(b) * sin(a_p);

dy_tm_ap =  cos(b) * cos(a_p);

dz_tm_ap =  sin(b);

G_tm_ap(i,j) = line_integral_from_source(x_tm, y_tm, z0, dx_tm_ap, dy_tm_ap, dz_tm_ap);


% ---- θ-dθ, α=θ-dα ----

dx_tm_am = -cos(b) * sin(a_m);

dy_tm_am =  cos(b) * cos(a_m);

dz_tm_am =  sin(b);

G_tm_am(i,j) = line_integral_from_source(x_tm, y_tm, z0, dx_tm_am, dy_tm_am, dz_tm_am);


% ---- s+ds ----

x_sp = (s0+ds) * cos(t);

y_sp = (s0+ds) * sin(t);

G_sp(i,j) = line_integral_from_source(x_sp, y_sp, z0, dx, dy, dz);


% ---- s-ds ----

x_sm = (s0-ds) * cos(t);

y_sm = (s0-ds) * sin(t);

G_sm(i,j) = line_integral_from_source(x_sm, y_sm, z0, dx, dy, dz);


% ---- s+ds, β+dbeta ----

b_p = b + dbeta;

dx_bp = -cos(b_p) * sin(t);

dy_bp =  cos(b_p) * cos(t);

dz_bp =  sin(b_p);
```

```
        G_sp_bp(i,j) = line_integral_from_source(x_sp, y_sp, z0, dx_bp, dy_bp, dz_bp);


        % ---- s+ds, β-dbeta ----
        b_m = b - dbeta;
        dx_bm = -cos(b_m) * sin(t);
        dy_bm =  cos(b_m) * cos(t);
        dz_bm =  sin(b_m);
        G_sp_bm(i,j) = line_integral_from_source(x_sp, y_sp, z0, dx_bm, dy_bm, dz_bm);


        % ---- s-ds, β+dbeta ----
        G_sm_bp(i,j) = line_integral_from_source(x_sm, y_sm, z0, dx_bp, dy_bp, dz_bp);


        % ---- s-ds, β-dbeta ----
        G_sm_bm(i,j) = line_integral_from_source(x_sm, y_sm, z0, dx_bm, dy_bm, dz_bm);
    end
end


%% 3. Compute partial derivatives (central differences)
fprintf('Computing partial derivatives...\n');


% ∂²G/∂θ∂α (at α=θ)
d2G_dtheta_dalpha = (G_tp_ap - G_tp_am - G_tm_ap + G_tm_am) / (4 * dtheta * dalpha);


% ∂G/∂s
dG_ds = (G_sp - G_sm) / (2 * ds);


% ∂²G/∂s∂β
```

```
dGdbeta_sp = (G_sp_bp - G_sp_bm) / (2 * dbeta);

dGdbeta_sm = (G_sm_bp - G_sm_bm) / (2 * dbeta);

d2G_dsdbeta = (dGdbeta_sp - dGdbeta_sm) / (2 * ds);


%% 4. Verify the equation

fprintf('\nVerification results (only points where the ray intersects the object):\n');

fprintf('--------------------------------------------------------------------------\n');

fprintf('   theta      beta       G0          LHS          RHS          error\n');

fprintf('--------------------------------------------------------------------------\n');


total_err = 0;

count = 0;

for i = 1:n_beta

    b = beta(i);

    for j = 1:n_theta

        if G0(i,j) < 1e-6

            continue;   % skip points where the ray barely intersects the object

        end


        % Left-hand side: ∂²G/∂θ∂α

        LHS = d2G_dtheta_dalpha(i,j);


        % Right-hand side: s cos²β ∂G/∂s + s cosβ sinβ ∂²G/∂s∂β

        RHS = s0 * cos(b)^2 * dG_ds(i,j) + s0 * cos(b) * sin(b) * d2G_dsdbeta(i,j);


        err = abs(LHS - RHS);

        fprintf(' %8.3f  %8.3f  %8.4f  %12.4e  %12.4e  %10.4e\n', ...
```

```
            theta(j), b, G0(i,j), LHS, RHS, err);


        total_err = total_err + err;
        count = count + 1;
    end
end
fprintf('---------------------------------------------------------------------------\n');
if count > 0
    fprintf('Number of valid points: %d\n', count);
    fprintf('Mean absolute error: %e\n', total_err / count);
else
    fprintf('Warning: No ray intersects the object. Check phantom or parameter range.\n');
end


%% Auxiliary function: line integral through Shepp-Logan phantom from a given source
function u = line_integral_from_source(x0, y0, z0, dx, dy, dz)
    % Shepp-Logan phantom parameters (center, semi-axes, density)
    ellipsoids = [
        0.3,   0.1,    0,      0.69,   0.92,   0.9,    2.0;
        0,     0,      0,      0.6624, 0.874,  0.88,  -0.98;
       -0.22,  0,     -0.25,   0.41,   0.16,   0.21,  -0.2;
        0.22,  0,     -0.25,   0.31,   0.11,   0.22,  -0.2;
        0,     0.35,  -0.25,   0.21,   0.25,   0.5,    0.2;
        0,     0.1,    0.625,  0.046,  0.046,  0.046,  0.2;
       -0.08, -0.605,  0,      0.046,  0.023,  0.02,   0.1;
        0.06, -0.605,  0,      0.046,  0.023,  0.02,   0.1;
        0,    -0.1,    0.625,  0.056,  0.04,   0.1,    0.2;
```

```
    0.06, -0.105,  0.625,  0.056,  0.056,  0.1,   -0.2;
];

u = 0;
tol = 1e-8;
for k = 1:size(ellipsoids, 1)
    xc = ellipsoids(k,1);
    yc = ellipsoids(k,2);
    zc = ellipsoids(k,3);
    a  = ellipsoids(k,4);
    b  = ellipsoids(k,5);
    c  = ellipsoids(k,6);
    density = ellipsoids(k,7);

    A = dx^2/a^2 + dy^2/b^2 + dz^2/c^2;
    if abs(A) < tol
        continue;
    end
    B = 2 * ( dx*(x0-xc)/a^2 + dy*(y0-yc)/b^2 + dz*(z0-zc)/c^2 );
    C = (x0-xc)^2/a^2 + (y0-yc)^2/b^2 + (z0-zc)^2/c^2 - 1;
    disc = B^2 - 4*A*C;
    if disc > tol
        sqrt_disc = sqrt(disc);
        t1 = (-B - sqrt_disc) / (2*A);
        t2 = (-B + sqrt_disc) / (2*A);
        u = u + density * (t2 - t1);
    end
```

```
    end
end
```

## Appendix G: MATLAB Code for Validating Formula (6.4)

```
%% Verify Eq. (6.4): -cosβ sinβ ∂G/∂s + cos²β ∂²G/∂s∂β = - ∂²G/∂z0∂α (α=θ)
% Using Shepp-Logan phantom for line integral computation

close all; clear all; clc;

%% 1. Parameter settings
s0 = 0.5;       % Radial distance of source point (baseline)
ds = 0.01;      % Step size for ∂/∂s
z0 = 0.1;       % Axial position of source point (baseline)
dz0 = 0.01;     % Step size for ∂/∂z0
rho = 1;        % Ray length (homogeneity of degree -1, fixed to 1)
theta = linspace(-pi/4, pi/4, 41);   % Source azimuth angle (also ray azimuth α)
beta = linspace(-pi/4, pi/4, 41);    % Ray tilt angle
dtheta = theta(2) - theta(1);
dbeta = beta(2) - beta(1);
n_theta = length(theta);
n_beta = length(beta);

fprintf('Verifying Eq. (6.4):\n');
fprintf('s0 = %.3f, ds = %.3f, z0 = %.2f, dz0 = %.4f\n', s0, ds, z0, dz0);
fprintf('Number of theta points: %d, number of beta points: %d\n', n_theta, n_beta);

%% 2. Compute required G values
% Need to compute G at the following positions:
```

```
% G0      : (s0, theta, z0)
% G_sp    : (s0+ds, theta, z0)
% G_sm    : (s0-ds, theta, z0)
% G_sp_bp : (s0+ds, theta, z0, beta+dbeta)
% G_sp_bm : (s0+ds, theta, z0, beta-dbeta)
% G_sm_bp : (s0-ds, theta, z0, beta+dbeta)
% G_sm_bm : (s0-ds, theta, z0, beta-dbeta)
% G_zp    : (s0, theta, z0+dz0)
% G_zm    : (s0, theta, z0-dz0)
% G_zp_tp : (s0, theta+dtheta, z0+dz0)  for ∂²G/∂z0∂α (α=θ)
% G_zp_tm : (s0, theta-dtheta, z0+dz0)
% G_zm_tp : (s0, theta+dtheta, z0-dz0)
% G_zm_tm : (s0, theta-dtheta, z0-dz0)


% Preallocate memory
G0 = zeros(n_beta, n_theta);
G_sp = zeros(n_beta, n_theta);
G_sm = zeros(n_beta, n_theta);
G_sp_bp = zeros(n_beta, n_theta);
G_sp_bm = zeros(n_beta, n_theta);
G_sm_bp = zeros(n_beta, n_theta);
G_sm_bm = zeros(n_beta, n_theta);
G_zp = zeros(n_beta, n_theta);
G_zm = zeros(n_beta, n_theta);
G_zp_tp = zeros(n_beta, n_theta);
G_zp_tm = zeros(n_beta, n_theta);
G_zm_tp = zeros(n_beta, n_theta);
```

```
G_zm_tm = zeros(n_beta, n_theta);

fprintf('Computing required G values...\n');

for i = 1:n_beta
    b = beta(i);
    for j = 1:n_theta
        t = theta(j);

        % Source point coordinates
        x0 = s0 * cos(t);
        y0 = s0 * sin(t);

        % Ray direction (α=θ, ρ=1)
        dx = -cos(b) * sin(t);
        dy =  cos(b) * cos(t);
        dz =  sin(b);

        % Baseline G(s0,θ,z0)
        G0(i,j) = line_integral_from_source(x0, y0, z0, dx, dy, dz);

        % G(s0+ds,θ,z0)
        x_sp = (s0+ds) * cos(t);
        y_sp = (s0+ds) * sin(t);
        G_sp(i,j) = line_integral_from_source(x_sp, y_sp, z0, dx, dy, dz);

        % G(s0-ds,θ,z0)
```

```
x_sm = (s0-ds) * cos(t);

y_sm = (s0-ds) * sin(t);

G_sm(i,j) = line_integral_from_source(x_sm, y_sm, z0, dx, dy, dz);


% G(s0+ds,θ,z0,β+Δβ)

b_p = b + dbeta;

dx_bp = -cos(b_p) * sin(t);

dy_bp =  cos(b_p) * cos(t);

dz_bp =  sin(b_p);

G_sp_bp(i,j) = line_integral_from_source(x_sp, y_sp, z0, dx_bp, dy_bp, dz_bp);


% G(s0+ds,θ,z0,β-Δβ)

b_m = b - dbeta;

dx_bm = -cos(b_m) * sin(t);

dy_bm =  cos(b_m) * cos(t);

dz_bm =  sin(b_m);

G_sp_bm(i,j) = line_integral_from_source(x_sp, y_sp, z0, dx_bm, dy_bm, dz_bm);


% G(s0-ds,θ,z0,β+Δβ)

G_sm_bp(i,j) = line_integral_from_source(x_sm, y_sm, z0, dx_bp, dy_bp, dz_bp);


% G(s0-ds,θ,z0,β-Δβ)

G_sm_bm(i,j) = line_integral_from_source(x_sm, y_sm, z0, dx_bm, dy_bm, dz_bm);


% G(s0,θ,z0+Δz0)

G_zp(i,j) = line_integral_from_source(x0, y0, z0+dz0, dx, dy, dz);
```

```
        % G(s0,θ,z0-Δz0)
        G_zm(i,j) = line_integral_from_source(x0, y0, z0-dz0, dx, dy, dz);

        % Four points for ∂²G/∂z0∂α
        % Note: α = θ, so changing θ corresponds to changing α
        t_p = t + dtheta;
        t_m = t - dtheta;

        % Ray directions must also change accordingly (since α = θ)
        dx_tp = -cos(b) * sin(t_p);
        dy_tp =  cos(b) * cos(t_p);
        dz_tp =  sin(b);
        dx_tm = -cos(b) * sin(t_m);
        dy_tm =  cos(b) * cos(t_m);
        dz_tm =  sin(b);

        % G(z0+Δz0, θ+Δθ)
        G_zp_tp(i,j) = line_integral_from_source(x0, y0, z0+dz0, dx_tp, dy_tp, dz_tp);
        % G(z0+Δz0, θ-Δθ)
        G_zp_tm(i,j) = line_integral_from_source(x0, y0, z0+dz0, dx_tm, dy_tm, dz_tm);
        % G(z0-Δz0, θ+Δθ)
        G_zm_tp(i,j) = line_integral_from_source(x0, y0, z0-dz0, dx_tp, dy_tp, dz_tp);
        % G(z0-Δz0, θ-Δθ)
        G_zm_tm(i,j) = line_integral_from_source(x0, y0, z0-dz0, dx_tm, dy_tm, dz_tm);
    end
end
```

```
%% 3. Compute partial derivatives (central difference)
fprintf('Computing partial derivatives...\n');

% ∂G/∂s
dG_ds = (G_sp - G_sm) / (2*ds);

% ∂²G/∂s∂β
% First compute ∂G/∂β at s+ds and s-ds
dGdbeta_sp = (G_sp_bp - G_sp_bm) / (2*dbeta);
dGdbeta_sm = (G_sm_bp - G_sm_bm) / (2*dbeta);
d2G_dsdbeta = (dGdbeta_sp - dGdbeta_sm) / (2*ds);

% ∂²G/∂z0∂α (note ∂α = ∂θ)
% First compute ∂G/∂z0 at θ+Δθ and θ-Δθ
dGdz0_tp = (G_zp_tp - G_zm_tp) / (2*dz0);
dGdz0_tm = (G_zp_tm - G_zm_tm) / (2*dz0);
% Then differentiate with respect to θ
d2G_dz0dalpha = zeros(n_beta, n_theta);
for i = 1:n_beta
   for j = 1:n_theta
      d2G_dz0dalpha(i,j) = (dGdz0_tp(i,j) - dGdz0_tm(i,j)) / (2*dtheta);
   end
end

%% 4. Verify the equation
fprintf('\nVerification results (only points where ray passes through object):\n');
fprintf('------------------------------------------------------------------\n');
```

```
fprintf(' theta   beta      G0          LHS            RHS          Error\n');
fprintf('-------------------------------------------------------------------\n');

total_err = 0;
count = 0;

for i = 1:n_beta
    b = beta(i);
    for j = 1:n_theta
        if G0(i,j) < 1e-6   % Skip points where ray barely passes through object
            continue;
        end

        % Left side: -cosβ sinβ * dG/ds + cos²β * d²G/dsdbeta
        LHS = -cos(b)*sin(b) * dG_ds(i,j) + (cos(b)^2) * d2G_dsdbeta(i,j);

        % Right side: - d²G/dz0dalpha
        RHS = - d2G_dz0dalpha(i,j);

        err = abs(LHS - RHS);

        fprintf(' %8.3f %8.3f %8.4f %12.4e %12.4e %10.4e\n', ...
                theta(j), b, G0(i,j), LHS, RHS, err);

        total_err = total_err + err;
        count = count + 1;
    end
```

```
end


fprintf('------------------------------------------------------------------\n');

if count > 0

    fprintf('Number of valid points: %d\n', count);

    fprintf('Mean absolute error: %e\n', total_err / count);

else

    fprintf('Warning: No valid points where ray passes through object. Check phantom or parameter range.\n');

end


%% Helper function: line integral through Shepp-Logan phantom from an arbitrary source point

function u = line_integral_from_source(x0, y0, z0, dx, dy, dz)

    % Shepp-Logan phantom parameters: [center x, center y, center z, semi-axis a, semi-axis b, semi-axis c, density]

    ellipsoids = [

        0.3,   0.1,   0,    0.69,  0.92,  0.9,   2.0;

        0,     0,     0,    0.6624,0.874, 0.88, -0.98;

       -0.22,  0,    -0.25, 0.41,  0.16,  0.21, -0.2;

        0.22,  0,    -0.25, 0.31,  0.11,  0.22, -0.2;

        0,     0.35, -0.25, 0.21,  0.25,  0.5,   0.2;

        0,     0.1,   0.625, 0.046, 0.046, 0.046, 0.2;

       -0.08, -0.605, 0,    0.046, 0.023, 0.02,  0.1;

        0.06, -0.605, 0,    0.046, 0.023, 0.02,  0.1;

        0,    -0.1,   0.625, 0.056, 0.04,  0.1,   0.2;

        0.06, -0.105, 0.625, 0.056, 0.056, 0.1,  -0.2;

    ];
```

```
u = 0;
tol = 1e-8;

for k = 1:size(ellipsoids, 1)
    xc = ellipsoids(k,1);
    yc = ellipsoids(k,2);
    zc = ellipsoids(k,3);
    a  = ellipsoids(k,4);
    b  = ellipsoids(k,5);
    c  = ellipsoids(k,6);
    density = ellipsoids(k,7);

    A = dx^2/a^2 + dy^2/b^2 + dz^2/c^2;
    if abs(A) < tol
        continue;
    end

    B = 2 * ( dx*(x0-xc)/a^2 + dy*(y0-yc)/b^2 + dz*(z0-zc)/c^2 );
    C = (x0-xc)^2/a^2 + (y0-yc)^2/b^2 + (z0-zc)^2/c^2 - 1;
    disc = B^2 - 4*A*C;

    if disc > tol
        sqrt_disc = sqrt(disc);
        t1 = (-B - sqrt_disc) / (2*A);
        t2 = (-B + sqrt_disc) / (2*A);
        % Only consider the forward direction t >= 0
        t_start = t1;
```

```
            t_end   = t2;
            u = u + density * (t_end - t_start);
        end
    end
end
```

## Appendix G: MATLAB Code for Validating Formula (6.5)

```
%% Verify Eq. (6.5): G_{θβ} - tanβ * G_θ = - s * tanβ * G_{z0β} - s * G_{z0} (α=θ)
% Using Shepp-Logan phantom for line integral computation


close all; clear all; clc;


%% 1. Parameter settings
s = 0.5;        % Radial distance of source point (non-zero)
z0 = 0.1;       % Axial position of source point
dz0 = 0.01;     % Step size for computing derivatives with respect to z0
theta = linspace(-pi/4, pi/4, 41);   % Source azimuth angles
beta = linspace(-pi/4, pi/4, 41);    % Ray tilt angles
dtheta = theta(2) - theta(1);
dbeta = beta(2) - beta(1);
n_theta = length(theta);
n_beta = length(beta);


fprintf('s = %.2f, dz0 = %.3f\n', s, dz0);
fprintf('theta range: [%.3f, %.3f]\n', theta(1), theta(end));
fprintf('beta range: [%.3f, %.3f]\n', beta(1), beta(end));


%% 2. Compute G(theta, beta) at z0 = 0 and z0 = ±dz0
```

```
G0 = zeros(n_beta, n_theta);

Gp = zeros(n_beta, n_theta);

Gm = zeros(n_beta, n_theta);


fprintf('Computing G values (z0 = 0, ±%.3f)...\n', dz0);


for i = 1:n_beta

    b = beta(i);

    for j = 1:n_theta

        t = theta(j);


        % Source point coordinates (s, θ, z0)

        xs0 = s * cos(t);

        ys0 = s * sin(t);

        zs0 = z0;

        zsp = z0 + dz0;

        zsm = z0 - dz0;


        % Ray direction (ρ=1, α=θ)

        dx = -cos(b) * sin(t);

        dy =  cos(b) * cos(t);

        dz =  sin(b);

        % Direction is already normalized (since cos²β+sin²β=1)


        % Compute line integrals

        G0(i,j) = line_integral_from_source(xs0, ys0, zs0, dx, dy, dz);

        Gp(i,j) = line_integral_from_source(xs0, ys0, zsp, dx, dy, dz);
```

```
        Gm(i,j) = line_integral_from_source(xs0, ys0, zsm, dx, dy, dz);
    end
end

%% 3. Compute partial derivatives for each matrix (three-point central difference)
% Compute G_theta and G_theta_beta for each matrix
G0_theta = zeros(n_beta, n_theta);
G0_theta_beta = zeros(n_beta, n_theta);
Gp_theta = zeros(n_beta, n_theta);
Gp_theta_beta = zeros(n_beta, n_theta);
Gm_theta = zeros(n_beta, n_theta);
Gm_theta_beta = zeros(n_beta, n_theta);

for i = 1:n_beta
    b = beta(i);
    for j = 1:n_theta
        t = theta(j);

        % Source point coordinates for shifted theta
        xs0 = s * cos(t);
        ys0 = s * sin(t);
        xs0_tp1 = s * cos(t + dtheta);
        ys0_tp1 = s * sin(t + dtheta);
        xs0_tm1 = s * cos(t - dtheta);
        ys0_tm1 = s * sin(t - dtheta);

        zs0 = z0;
```

```
        zsp = z0 + dz0;
        zsm = z0 - dz0;

        % Ray direction (ρ=1, α=θ) at the baseline theta
        dx = -cos(b) * sin(t);
        dy =  cos(b) * cos(t);
        dz =  sin(b);

        % Compute line integrals at shifted theta (using same ray direction as baseline)
        G0_i_jp1 = line_integral_from_source(xs0_tp1, ys0_tp1, zs0, dx, dy, dz);
        G0_i_jm1 = line_integral_from_source(xs0_tm1, ys0_tm1, zs0, dx, dy, dz);
        Gp_i_jp1 = line_integral_from_source(xs0_tp1, ys0_tp1, zsp, dx, dy, dz);
        Gp_i_jm1 = line_integral_from_source(xs0_tm1, ys0_tm1, zsp, dx, dy, dz);
        Gm_i_jp1 = line_integral_from_source(xs0_tp1, ys0_tp1, zsm, dx, dy, dz);
        Gm_i_jm1 = line_integral_from_source(xs0_tm1, ys0_tm1, zsm, dx, dy, dz);

        G0_theta(i,j) = (G0_i_jp1 - G0_i_jm1) / (2 * dtheta);
        Gp_theta(i,j) = (Gp_i_jp1 - Gp_i_jm1) / (2 * dtheta);
        Gm_theta(i,j) = (Gm_i_jp1 - Gm_i_jm1) / (2 * dtheta);
    end
end

% Compute G_theta_beta (mixed derivative) for G0
for i = 1:n_beta
    b = beta(i);
    for j = 1:n_theta
        t = theta(j);
```

```
% Source point coordinates for shifted theta
xs0 = s * cos(t);
ys0 = s * sin(t);
xs0_tp1 = s * cos(t + dtheta);
ys0_tp1 = s * sin(t + dtheta);
xs0_tm1 = s * cos(t - dtheta);
ys0_tm1 = s * sin(t - dtheta);

zs0 = z0;

% Ray directions for shifted beta
dx_bp1 = -cos(b + dbeta) * sin(t);
dy_bp1 =  cos(b + dbeta) * cos(t);
dz_bp1 =  sin(b + dbeta);
dx_bm1 = -cos(b - dbeta) * sin(t);
dy_bm1 =  cos(b - dbeta) * cos(t);
dz_bm1 =  sin(b - dbeta);

% Compute line integrals at four corners (theta±, beta±)
G0_ip1_jp1 = line_integral_from_source(xs0_tp1, ys0_tp1, zs0, dx_bp1, dy_bp1, dz_bp1);
G0_ip1_jm1 = line_integral_from_source(xs0_tm1, ys0_tm1, zs0, dx_bp1, dy_bp1, dz_bp1);
G0_im1_jp1 = line_integral_from_source(xs0_tp1, ys0_tp1, zs0, dx_bm1, dy_bm1, dz_bm1);
G0_im1_jm1 = line_integral_from_source(xs0_tm1, ys0_tm1, zs0, dx_bm1, dy_bm1, dz_bm1);

% Central difference for mixed derivative: ∂²G/∂θ∂β
G0_theta_beta(i,j) = (G0_ip1_jp1 - G0_ip1_jm1 - G0_im1_jp1 + G0_im1_jm1) / (2 * dtheta) / (2
* dbeta);
```

```
    end
end

%% 4. Compute derivatives with respect to z0 (central difference)
G_z0 = (Gp - Gm) / (2 * dz0);

G_z0_beta = zeros(n_beta, n_theta);
for i = 1:n_beta
    b = beta(i);
    for j = 1:n_theta
        t = theta(j);

        % Source point coordinates
        xs0 = s * cos(t);
        ys0 = s * sin(t);
        zs0 = z0;
        zsp = z0 + dz0;
        zsm = z0 - dz0;

        % Ray directions for shifted beta
        dx_bp1 = -cos(b + dbeta) * sin(t);
        dy_bp1 =  cos(b + dbeta) * cos(t);
        dz_bp1 =  sin(b + dbeta);
        dx_bm1 = -cos(b - dbeta) * sin(t);
        dy_bm1 =  cos(b - dbeta) * cos(t);
        dz_bm1 =  sin(b - dbeta);
```

```
        % Compute line integrals at (z0±dz0, beta±dbeta)
        Gp_ip1_j = line_integral_from_source(xs0, ys0, zsp, dx_bp1, dy_bp1, dz_bp1);
        Gp_im1_j = line_integral_from_source(xs0, ys0, zsp, dx_bm1, dy_bm1, dz_bm1);
        Gm_ip1_j = line_integral_from_source(xs0, ys0, zsm, dx_bp1, dy_bp1, dz_bp1);
        Gm_im1_j = line_integral_from_source(xs0, ys0, zsm, dx_bm1, dy_bm1, dz_bm1);

        % Mixed derivative: ∂²G/∂z0∂β
        G_z0_beta(i,j) = (Gp_ip1_j - Gp_im1_j - Gm_ip1_j + Gm_im1_j) / (2 * dbeta) / (2 * dz0);
    end
end

%% 5. Verify equation (6.5)
fprintf('\nVerifying Eq. (6.5): G_{θβ} - tanβ * G_θ = - s * tanβ * G_{z0β} - s * G_{z0}\n');
fprintf('-----------------------------------------------------------------------------------------\n');
fprintf(' theta   beta      G0          LHS           RHS          Error\n');
fprintf('-----------------------------------------------------------------------------------------\n');

total_error = 0;
count = 0;

for i = 1:n_beta
    b = beta(i);
    for j = 1:n_theta
        t = theta(j);
        if G0(i,j) < 1e-6
            continue;   % Skip points where ray does not pass through object
        end
```

```
        lhs = G0_theta_beta(i,j) - tan(b) * G0_theta(i,j);
        rhs = -s * tan(b) * G_z0_beta(i,j) - s * G_z0(i,j);
        err = abs(lhs - rhs);

        fprintf(' %8.3f %8.3f %8.4f %8.4e %8.4e %8.4e\n', ...
            t, b, G0(i,j), lhs, rhs, err);

        total_error = total_error + err;
        count = count + 1;
    end
end

fprintf('-----------------------------------------------------------------------------------------\n');
if count > 0
    fprintf('Number of valid points: %d\n', count);
    fprintf('Mean absolute error: %e\n', total_error / count);
end

%% Helper function: line integral through Shepp-Logan phantom from an arbitrary source point
function u = line_integral_from_source(x0, y0, z0, dx, dy, dz)
    % Shepp-Logan phantom parameters: [center x, center y, center z, semi-axis a, semi-axis b, semi-axis c, density]
    ellipsoids = [
        0.3,   0.1,   0,    0.69,  0.92,  0.9,   2.0;
        0,     0,     0,    0.6624,0.874, 0.88, -0.98;
       -0.22,  0,    -0.25, 0.41,  0.16,  0.21, -0.2;
        0.22,  0,    -0.25, 0.31,  0.11,  0.22, -0.2;
```

```
    0,     0.35, -0.25, 0.21,  0.25,  0.5,   0.2;
    0,     0.1,   0.625, 0.046, 0.046, 0.046, 0.2;
   -0.08, -0.605, 0,    0.046, 0.023, 0.02,  0.1;
    0.06, -0.605, 0,    0.046, 0.023, 0.02,  0.1;
    0,    -0.1,   0.625, 0.056, 0.04,  0.1,   0.2;
    0.06, -0.105, 0.625, 0.056, 0.056, 0.1,  -0.2;
];

u = 0;
tol = 1e-8;

for k = 1:size(ellipsoids, 1)
    xc = ellipsoids(k,1);
    yc = ellipsoids(k,2);
    zc = ellipsoids(k,3);
    a  = ellipsoids(k,4);
    b  = ellipsoids(k,5);
    c  = ellipsoids(k,6);
    density = ellipsoids(k,7);

    A = dx^2/a^2 + dy^2/b^2 + dz^2/c^2;
    if abs(A) < tol
        continue;
    end

    B = 2 * ( dx*(x0-xc)/a^2 + dy*(y0-yc)/b^2 + dz*(z0-zc)/c^2 );
    C = (x0-xc)^2/a^2 + (y0-yc)^2/b^2 + (z0-zc)^2/c^2 - 1;
```

```
        disc = B^2 - 4*A*C;


        if disc > tol
            sqrt_disc = sqrt(disc);
            t1 = (-B - sqrt_disc) / (2*A);
            t2 = (-B + sqrt_disc) / (2*A);
            % Only consider t >= 0 (forward direction)
            t_start = t1;
            t_end   = t2;
            u = u + density * (t_end - t_start);
        end
    end
end
```